\documentclass[twocolumn,amsmath,amssymb,,showpacs,prb, preprintnumbers]
{revtex4}

\usepackage{graphicx,dcolumn,bm,pslatex}

\begin{document}
\preprint{APS/123-QED}
\title{Understanding complex magnetic order in disordered cobalt hydroxides through analysis of the local structure}

\author{James R. Neilson}
\email{neilson@lifesci.ucsb.edu}
\author{Daniel E. Morse}
\email{d_morse@lifesci.ucsb.edu}
\affiliation{
Biomolecular Science \& Engineering\\
University of California Santa Barbara, CA 93106}
\author{Brent C. Melot}
\author{Daniel P. Shoemaker}
\author{Joshua A. Kurzman}
\author{Ram Seshadri}
\affiliation{
Materials Department and Materials Research Laboratory\\ 
University of California Santa Barbara, CA 93106
}

\date{\today}

\begin{abstract}
In many ostensibly crystalline materials, unit-cell-based descriptions do not always capture the complete physics of the system due to disruption in long-range order.  In the series of cobalt hydroxides studied here, Co(OH)$_{2-x}$(Cl)$_x$(H$_2$O)$_{n}$, magnetic Bragg diffraction reveals a fully compensated N\'eel state, yet the materials show significant and open magnetization loops.  A detailed analysis of the local structure defines the aperiodic arrangement of cobalt coordination polyhedra. Representation of the structure as a combination of distinct polyhedral motifs explains the existence of locally uncompensated moments and provides a quantitative agreement with bulk magnetic measurements and magnetic Bragg diffraction. 
\pacs{75.25.-j, 64.60.Cn, 61.66.Fn}
\end{abstract}
\maketitle

\section{Introduction}

Complex magnetic ground states in crystalline materials frequently derive from the details 
of the underlying lattices. In the popular example of 
insulating compounds where the magnetic ions arrange themselves with three-fold symmetries 
(\emph{e.g.} triangular or pyrochlore lattices),  
geometric frustration of magnetic interactions prevents magnetic ordering or 
gives rise to exotic ground states.\cite{Ramirez_GeoFrust}  
Alternatively, the dilution of magnetic ions can yield complex magnetic 
order, since the translational periodicity of the lattice is necessarily interrupted.   

For example, in diluted uniaxial anisotropic antiferromagnets, 
such as  Co$_{1-x}$Zn$_{x}$F$_2$, 
local uncompensated moments result from random variations in the exchange 
interactions from the non-magnetic impurities.\cite{Fishman_1979}   
Upon zero-field cooling, neutron diffraction reveals a N\'eel state; however, the random fields 
decrease the diffracted peak intensity from disrupted long-range order (LRO) and field 
cooling completely destroys the LRO.\cite{PhysRevLett.48.438, PhysRevB.28.2602}  
Because these systems show no preference for chemical or short-range order within a 
disordered nuclear unit-cell,\cite{PhysRevB.28.2602} it is difficult to structurally 
distinguish the role of the impurities in establishing the random fields and 
microdomains.  

\begin{figure}
\includegraphics[width =2.5in]{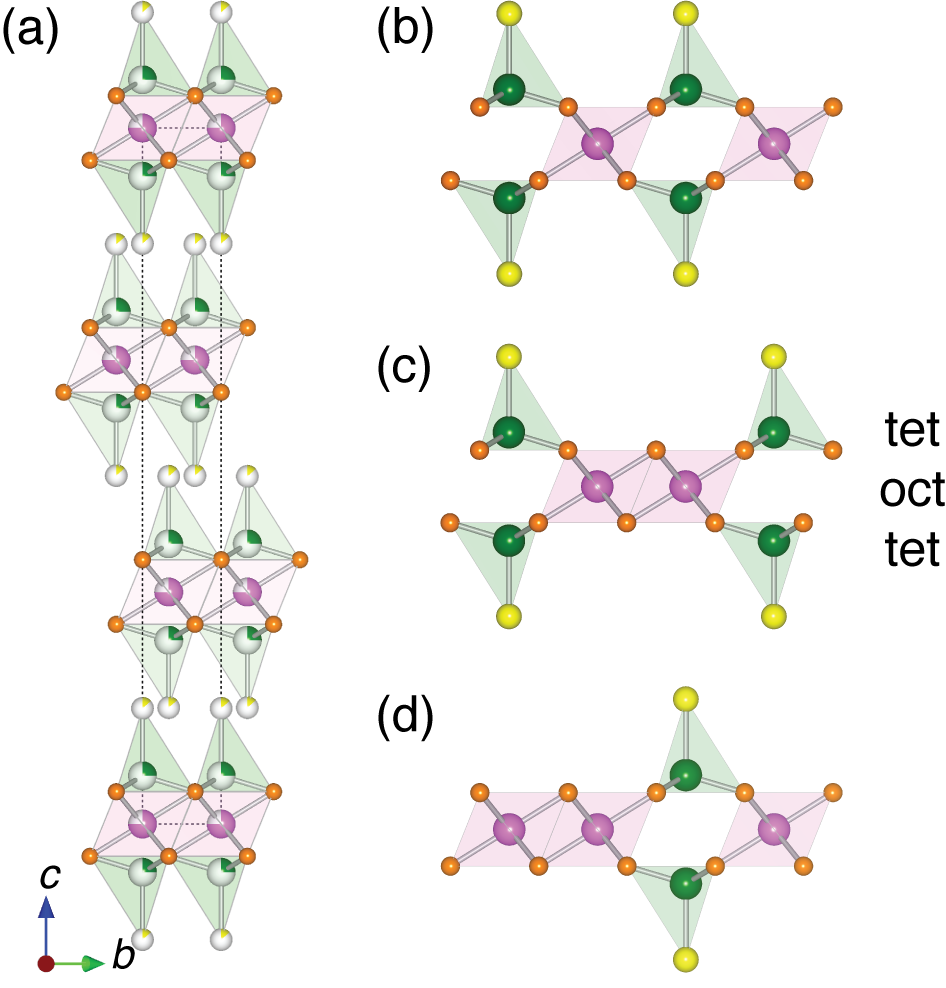}
\caption{\label{fig:structure} (Color online) (a) Average structure of
layered cobalt hydroxide chloride, Co(OH)$_{2-x}$(Cl)$_x$(H$_2$O)$_{n}$ (dashed 
line indicates the unit-cell) indicating the fractionally occupied (partially filled) metal sites.  
(b-d) Locally ordered distributions of capping tetrahedral Co$^{2+}$, with ratios of 2 Co$^{\text{tet}}$ per (b) 3 Co$^{\text{oct}}$, (c) 8 Co$^{\text{oct}}$, and (d) 15 Co$^{\text{oct}}$.  From Ref.\,\onlinecite{Neilson_ChemEurJ_2010}.  Pink spheres denote Co$^{\text{oct}}$, green Co$^{\text{tet}}$; oxygen is orange, chlorine yellow.  Hydrogen is omitted for clarity.   }
\end{figure}

The layered antiferromagnet, $\beta$-Co(OH)$_2$, contains only edge-sharing Co$^{2+}$ octahedra and shows field-induced magnetic behavior below 12.3 K.\cite{Takada_1966,Liu_APL_2008}  
In the $\alpha$-cobalt hydroxides studied here, the layers of edge-sharing octahedra are disrupted by stellating cobalt tetrahedra (Fig.\,\ref{fig:structure}).  These modified structures are often referred to as $\alpha$-Co(OH)$_2$.\cite{Oswald_Asper_1977}  The series of $\alpha$-cobalt hydroxide chlorides are described by the chemical formula 
Co(OH)$_{2-x}$(Cl)$_x$(H$_2$O)$_{n}$, where \emph{x}~= 0.2, 
0.3, and 0.4 assumes the tetrahedrally coordinated cobalt content, and $n$ = 1.3, 0.9,  and 0.3, respectively.\cite
{Neilson_IC_2009}  We refer to the $\beta$-Co(OH)$_2$ compounds as $x=0$.  

In a traditional unit-cell based description of the structure, all cobalt site occupancies 
($g$) are reduced ($g_{\text{oct}} = 1-g_{\text{tet}}$) and randomly averaged [Fig.
\ref{fig:structure}(a)].  
Therefore, a local description is needed to locate all of the atom positions and 
occupancies.  
From our extensive analysis of the X-ray pair distribution functions of these 
compounds,\cite{Neilson_ChemEurJ_2010} the distribution of cobalt polyhedra is 
best described by a random combination of structural motifs [Fig.\,\ref
{fig:structure}(b-d)].  We propose that chemical short-range order of the local structure generates complex magnetic behavior, as in diluted uniaxial anisotropic antiferromagnets.

Here we report magnetic susceptibility measurements to reveal uncompensated magnetization, yet magnetic neutron diffraction suggests a periodic N\'eel antiferromagnetic structure.  
We present a method of assembling local motifs in real space to microscopically 
reconcile a quantitative description of the random distribution of exchange interactions 
and the resulting microdomains.  

\section{Experimental}

The experimental samples with $x$ = 0.2, 0.3, and 0.4 were prepared and characterized as 
previously described\cite{Neilson_IC_2009, Neilson_ChemEurJ_2010} and material with 
$x=0.0$ was purchased from Alfa-Aesar.  Magnetic characterization was performed on powder samples embedded in paraffin wax (melted to $55^\circ$C) using a Quantum Design MPMS SQUID magnetometer.  

The specific heats of Co(OH)$_{2-x}$(Cl)$_{x}$(H$_2$O)$_n$ for $x =$ 0.2, 0.3, and 0.4 and the non-magnetic analogs Mg(OH)$_2$ and Zn(OH)$_{1.6}$Cl$_{0.4}$H$_2$O$_{0.2}$  were measured using a semi-adiabatic technique as implemented in a Quantum Design Physical Properties Measurement System.  All powder samples were mixed with Ag powder (50 wt\%) in an agate mortar and pestle until homogeneous.  Pellets of the mixed powder were pressed in a die (3 by 9 mm, $\sim$0.5 mm thickness) at 0.5 tons.  The silver provides mechanical stability to the pellet and improves thermal conductivity to prevent thermal gradients and increase the thermal relaxation.  Both stage with thermal grease and Ag calibrations were collected separately before measuring divided pellets (3 by 3 mm, approximately 10 mg), which were affixed to the sample stage using thermal grease.

A deuterated analog was synthesized from a D$_2$O solution for neutron experiments.    Neutron diffraction of  Co(OD)$_{1.6}$(Cl)$_{0.4}$(D$_2$O)$_{0.3}$ was recorded on the HIPD beamline at the Lujan Neutron Scattering Center at Los Alamos National Laboratory at both 6 K and 300 K.  Approximately 1 g of sample was packed into a vanadium canister.  To calculate and subtract sample absorption using \textsc{PDFgetN},\cite{PDFgetN}  scattering profiles of the empty vanadium container, the evacuated chamber, and a vanadium rod were also collected.  All other experimental details are described in the text where relevant.  Experimental uncertainty of reported values (\emph{e.g.} saturation magnetization, entropy) is determined from the sum in quadrature of fractional uncertainties, with the predominant source of error originating from determination of the measured sample mass.  

\section{Results and Discussion}

\subsection{Macroscopic Magnetization}

\begin{figure}
\centering
\includegraphics[width =3.0in]{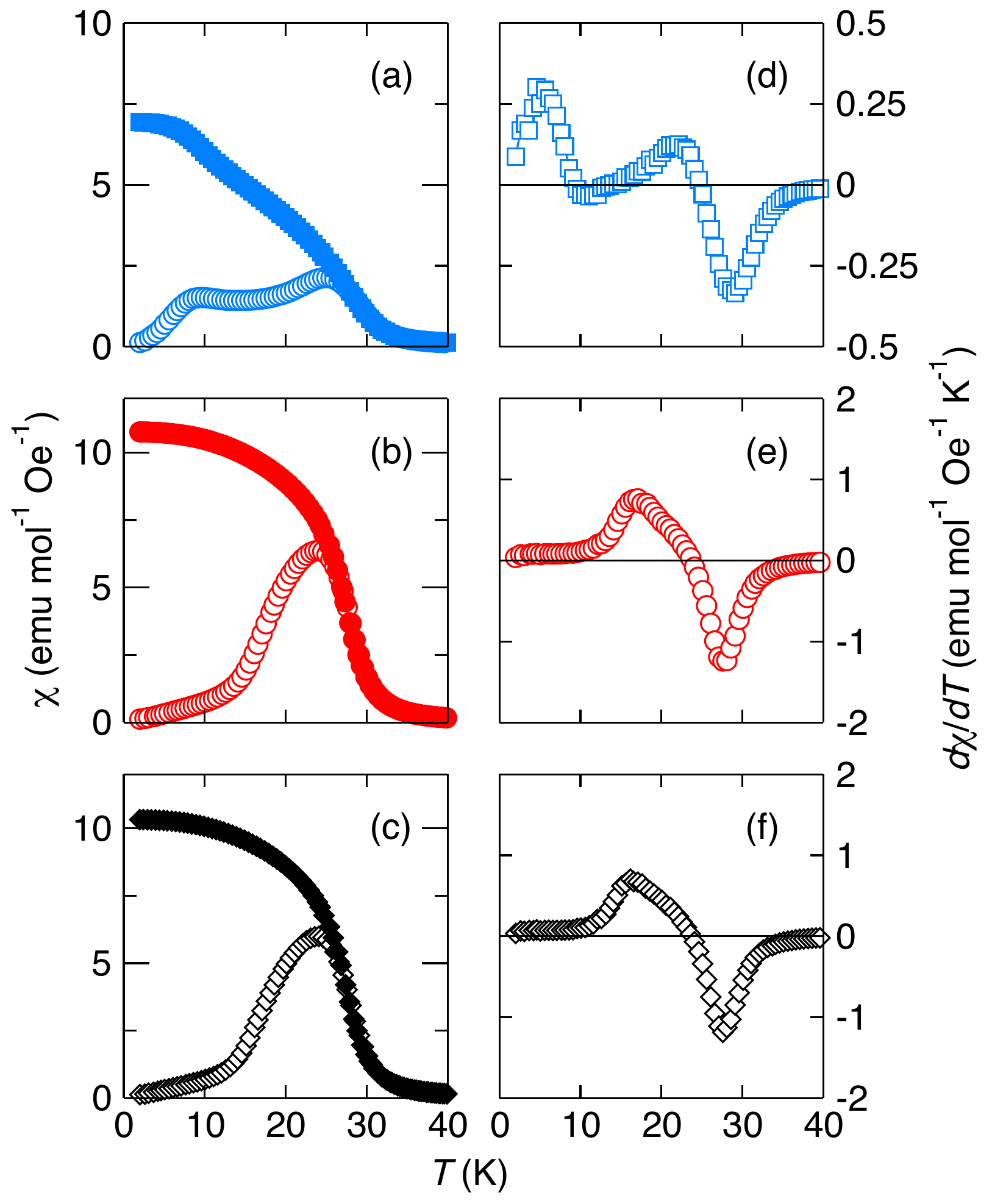}
\caption{\label{fig:suscept} (Color online) Field-cooled (FC, filled symbols) and zero-field cooled (ZFC, open symbols) molar dc susceptibility of Co(OH)$_{2-x}$(Cl)$_{x}$(H$_2$O)$_n$ for $x =$ (a) 0.2, (b) 0.3, and (c) 0.4 for  at low temperatures and the temperature derivative of the ZFC curves  ($d\chi/dT$) for (d) 0.2, (e) 0.3, and (f) 0.4 indicating two cusps for $x$ = 0.2 and an invariant $T_c$ near 24 K. Data acquired under a magnetic field of 100 Oe.  }
\end{figure}

\begin{figure}
\centering
\includegraphics[width =3in]{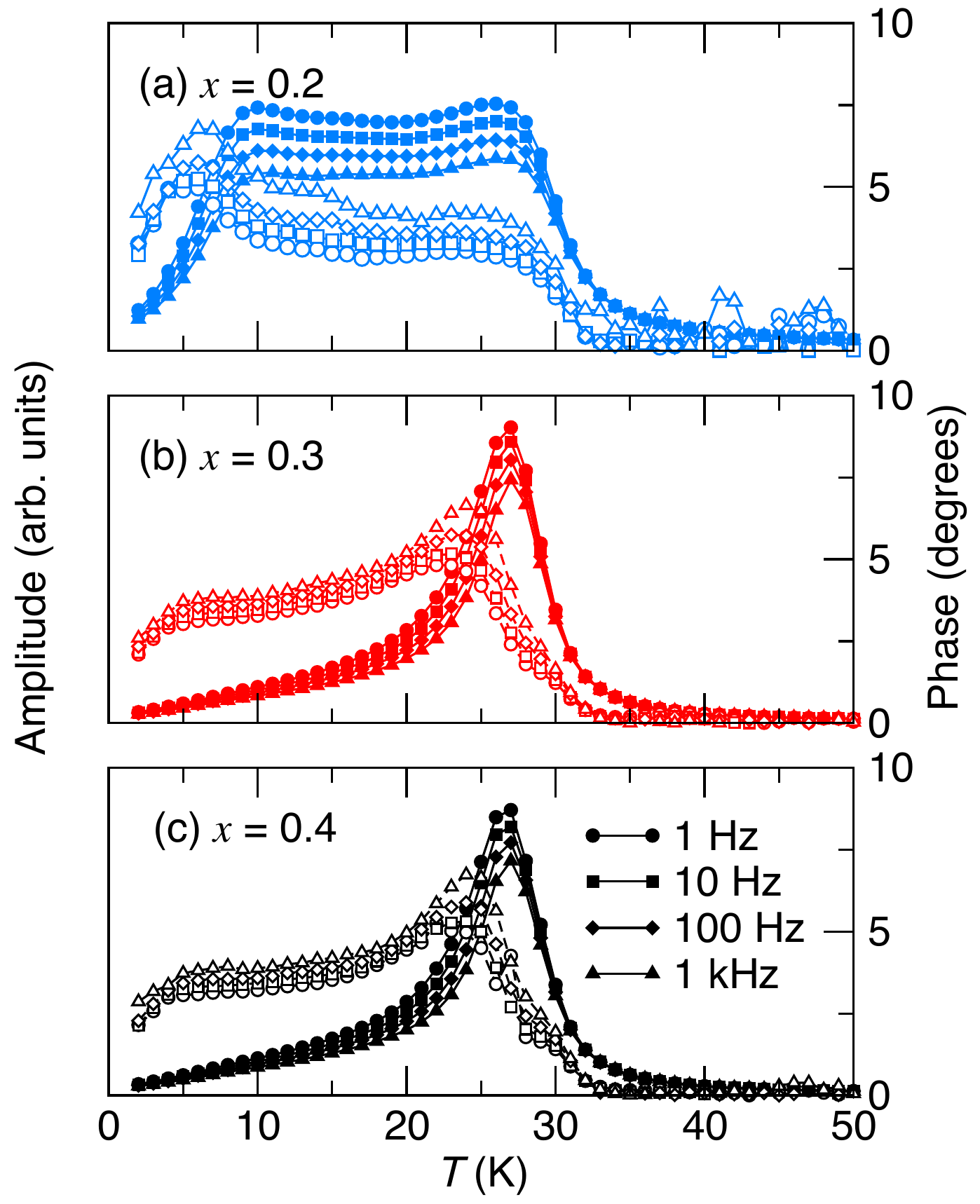}
\caption{\label{fig:ac} (Color online) Field-cooled frequency dependent amplitude (filled symbols) and phase (open symbols) from ac susceptibility measurements of Co(OH)$_{2-x}$(Cl)$_{x}$(H$_2$O)$_n$ for $x =$ (a) 0.2, (b) 0.3, and (c) 0.4 for low temperatures. Data were acquired with zero applied static magnetic field and an alternating field with an amplitude of 3 Oe. }
\end{figure}

The static magnetic susceptibilities, measured in zero-field and field-cooling conditions, show a nearly constant $T_c$ with composition, as determined by the tempearture at which $d\chi/dT = 0$ (Fig.\,\ref{fig:suscept}, Table\,\ref{tab:curie}).  The curves for $x=0.2$ show two cusps at approximately 28.4~K and 9.4~K [Fig.\,\ref{fig:suscept}(a,d)], suggestive of multiple ordering transitions or sublattice compensation points.  While the lower temperature maximum is obvious for $x=0.2$, there are subtle changes present in the $x=$ 0.3 and 0.4 compounds visible in the AC susceptibility, Fig\,\ref{fig:ac}.  All compounds show frequency dependent cusps in the phase near $T_c$, and again at $\sim$9~K.  The strong frequency dependence in the dynamic susceptibility is supportive of the freezing of glassy spins from different sublattices or microdomains.

\begin{table}
\caption{\label{tab:curie} Curie temperature determined from $d\chi/dT = 0$ and Curie constants obtained from a linear fit to the Curie-Weiss law from 200\,K - 300\,K.}
\begin{ruledtabular}
\begin{tabular}{c|ccc}
 $x$ & $T_c$ (K) & $\Theta_\text{CW}$ (K) & $C$ (emu mol$^{-1}$ Oe$^{-1}$) \\
 \hline
0.0 & 10.7 & 14.2 & 3.1\\
0.2 &  24.8, 9.4  & 31.2 & 2.0\\
0.3 &  23.7 & -20.0 & 2.8\\
0.4 &  23.7 & -16.9 & 2.7\\
\end{tabular}
\end{ruledtabular}
\label{default}
\end{table}%

\begin{figure}
\centering
\includegraphics[width=3in]{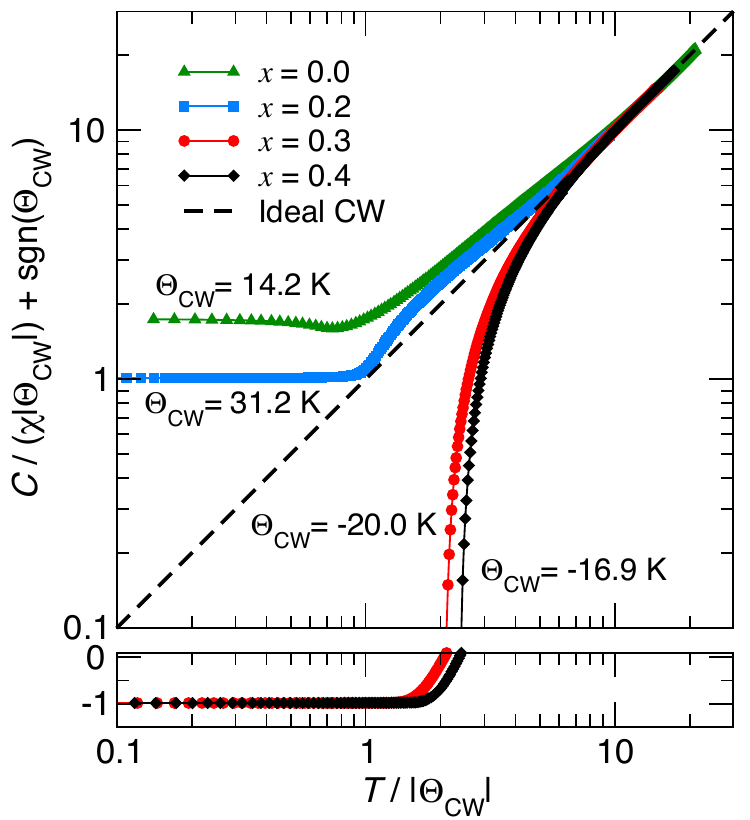}\\
\caption{\label{fig:normcw} (Color online) Temperature dependent scaled inverse 
susceptibility (field-cooled at 100 Oe) of Co(OH)$_{2-x}$(Cl)$_{x}$(H$_{2}$O)$_{n}$ for 
$x=$ 0.0, 
0.2, 0.3, and 0.4, illustrating positive
compensated and negative uncompensated deviations from ideal Curie-Weiss
paramagnetism.}
\end{figure}

From the paramagnetic regime (200\,K to 300\,K), the inverse susceptibility measured under field cooling is fit to the Curie-Weiss law, $1/\chi=(T-\Theta_{\text{CW}})/C$, to extract the constants $C$ and $\Theta_\text{CW}$ (Table\,\ref{tab:curie}).  From data scaled by $C$ and  $\Theta_\text{CW}$, 
\begin{equation}
\frac{C}{\chi|\Theta_{\text{CW}}|} + \text{sgn}(\Theta_{\text{CW}}) = \frac{T}{|\Theta_{\text{CW}}|},\label{eqn:cw}
\end{equation}
we compare the relative behavior of all four compounds (Figure\,\ref{fig:normcw}).  
The quality of fit to the Curie-Weiss law is noted by convergence of all data to the ideal Curie-Weiss relationship (dashed black line) at high temperatures in Figure\,\ref{fig:normcw}.  
All of the cobalt hydroxides show strong deflections from the Curie-Weiss law at $T/|\Theta_{\text{CW}}| = 1$, indicative of a magnetic phase transition.

All the compounds studied also show small deviations 
as  $T/|\Theta_{\text{CW}}| $ approaches unity on cooling.  When plotted according to Eqn.\,\ref{eqn:cw}, positive deflections from the ideal Curie-Weiss behavior are indicative of compensated interactions; negative deviations signify uncompensated interactions.\cite{Melot_2009_JPCM}  The all-octahedral $\beta$-Co(OH)$_2$ compound ($x=0$) shows compensated interactions when $T/|\Theta_{\text
{CW}}|  >1$, attributable to the longer-range ($\propto r^{-3}$) interlayer dipolar 
exchange.  This is also observed when the layers become decorated with tetrahedra 
($x=0.2$); however, uncompensated interactions emerge when $x>0.2$.  This crossover 
is reflected in $\Theta_{\text{CW}}$, which is positive (ferromagnetic) for $x\le0.2$, 
and negative (antiferromagnetic) as $x>0.2$.  As $x$ increases above  $x=0.2$, $|\Theta_
{\text{CW}}|$ decreases below $T_c$ (Table\,\ref{tab:curie}), suggesting a decrease in the interaction strength.  
From a mean-field perspective, we interpret this observation to indicate that the dominant interaction 
switches from intralayer ferromagnetic coupling when $x\le0.2$, as found in 
all-octahedrally coordinated cobalt(II) compounds,\cite{Rabu_IC_1993} to 
antiferromagnetic interactions between distinct polyhedra.   

\begin{figure}
\begin{center}
\includegraphics[width=3in]{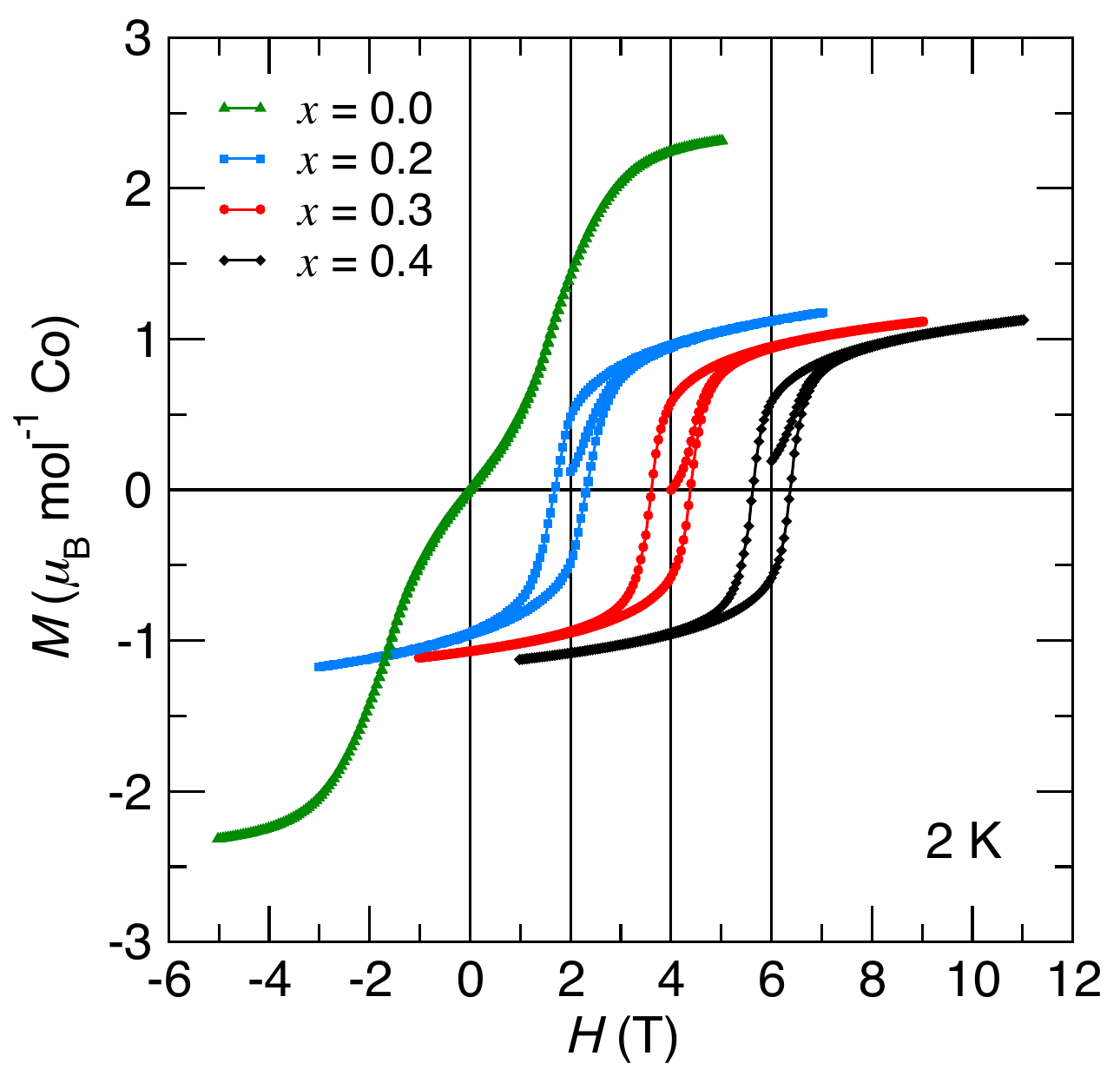}
\caption{\label{fig:hys}(Color online) Isothermal magnetization of Co(OH)$_{2-x}$(Cl)$_
x$(H$_2$O)$_n$ for $x=$ 0.0, 0.2, 0.3, and 0.4 at 2 K showing the composition 
independent saturation magnetization for $x > 0.0$.  The inflection point in the initial 
magnetization indicates a weak field induced transition. Each curve is offset by 2 T for 
clarity.  }
\end{center}
\end{figure}

\begin{table}
\caption{\label{tab:summary}Expected
saturation magnetization assuming  a two-sublattice N\'eel model,
saturation magnetization for an explicit treatment of the local structure,\cite{Neilson_ChemEurJ_2010} and the observed magnetization (2 K, 5 T) for each compound ($x = 0.0, 0.2, 0.3, 0.4$).  
Summary of the magnetic heat capacity measurements: lost entropy and temperature 
exponent at $T \ll T_c$.}
\begin{ruledtabular}
\begin{tabular}{c|ccc|cc}
 &  N\'eel & Local &Obs. & \multicolumn{2}{c}{Heat capacity}\\
 &  $M_{	sat}$& $M_{	sat}$ &$M_{	sat}$&$\Delta S_{\text{mag}}$  & $\beta$ \\
 $x$ &\multicolumn{3}{c|}{$(\mu_\text{B}$ mol$^{-1}$ Co)}& (J mol$^{-1}$ K$^
{-1}$)& 
($T^\beta$) \\
\hline
0.0 &  n/a  & n/a  &2.3$\pm$0.02 & 5.2$\pm$0.1  & 2.0\\
0.2 &  0.9  & 1.14$\pm$0.07 &1.2$\pm$0.02 & 3.7$\pm$0.1  & 1.9 \\
0.3 &  0.6  & 1.14$\pm$0.07 &1.1$\pm$0.02 & 3.6$\pm$0.1  & 2.0 \\
0.4 &  0.3  & 1.13$\pm$0.07 &1.1$\pm$0.02 & 3.4$\pm$0.1  & 2.3\\
\end{tabular}
\end{ruledtabular}
\label{default}
\end{table}%

The isothermal magnetization (2 K, Fig.\,\ref{fig:hys}) of $x=0$ is indicative of 
antiferromagnetic order with a spin-flop transition.\cite{Takada_1966,Rabu_IC_1993}  
The hysteresis loops open when $x>0$ with a composition independent saturation 
magnetization at 1.1-1.2 $\mu_\text{B}$ mol$^{-1}$ Co (Obs. $M_{sat}$ in Table\,\ref
{tab:summary}).   This independence contradicts previous studies\cite{Kurmoo_1999, Kurmoo:1999uq} on similar structure types where the N\'eel sublattices were assigned to different coordination polyhedral (Co$^{\text{oct}}$ \emph{vs} Co$^{\text{tet}}$).  Therefore, the expected magnetization is equal to 3/2(Co$^{\text{oct}}$ sites - Co$^{\text{tet}}$ sites).   With that spin arrangement, one calcualtes a composition dependent magnetization (Table\,\ref{tab:summary}, N\'eel $M_{sat}$).

\subsection{Low Temperature Heat Capacity}
 
\begin{figure}
\begin{center}
\includegraphics[width =3.in]{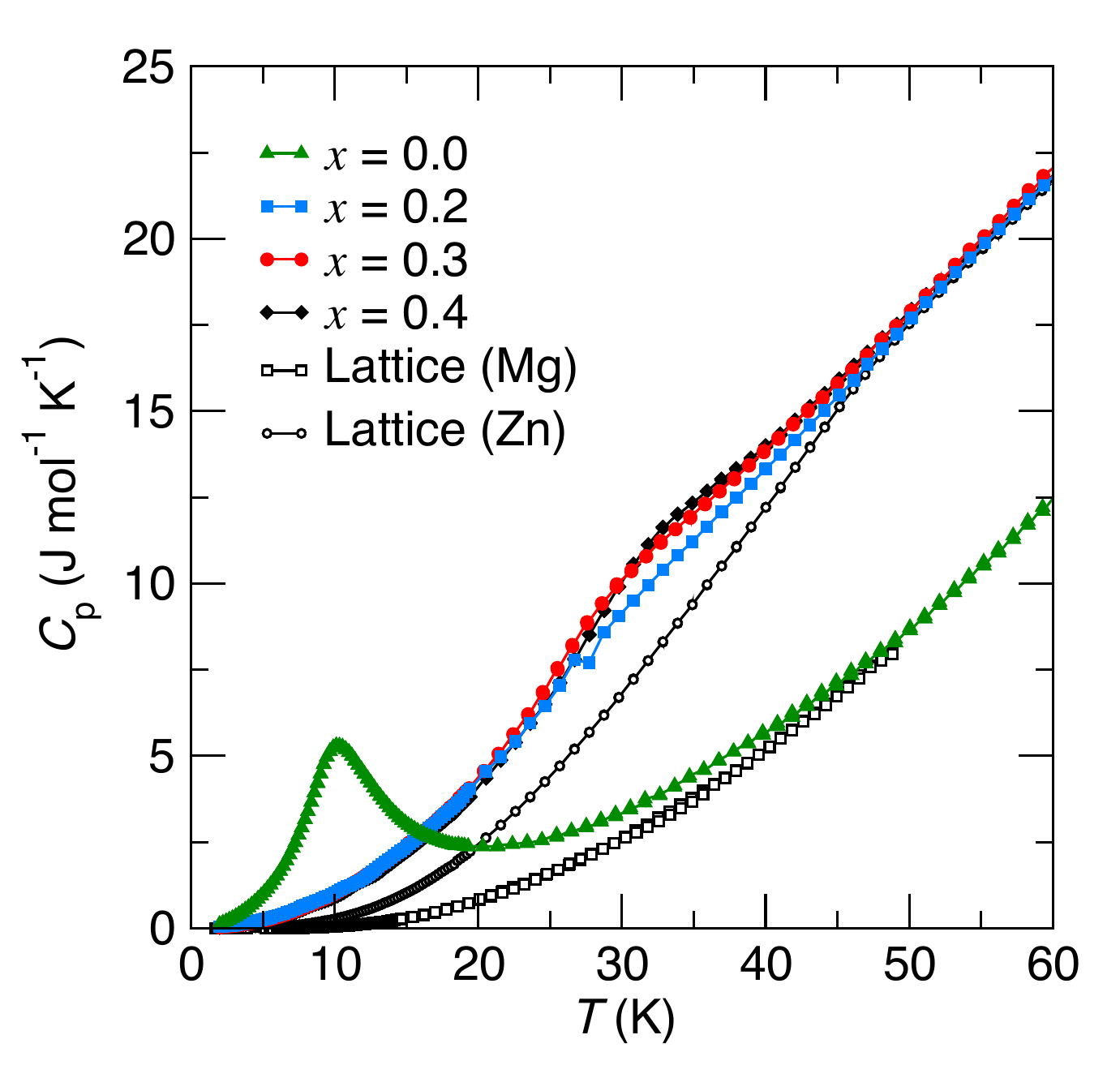}
\caption{\label{fig:hctotal}(Color online) Specific heat capacity measured at low temperature and zero applied field of Co(OH)$_{2-x}$(Cl)$_{x}$(H$_2$O)$_n$ for $x =$ (a) 0.2, (b) 0.3, and (c) 0.4 and the non-magnetic analogs Mg(OH)$_2$ and Zn(OH)$_{1.6}$Cl$_{0.4}$H$_2$O,  after accounting for the differing oscillator masses of the non-magnetic compounds. }
\end{center}
\end{figure}

\begin{figure}
\begin{center}
\includegraphics[width=3in]{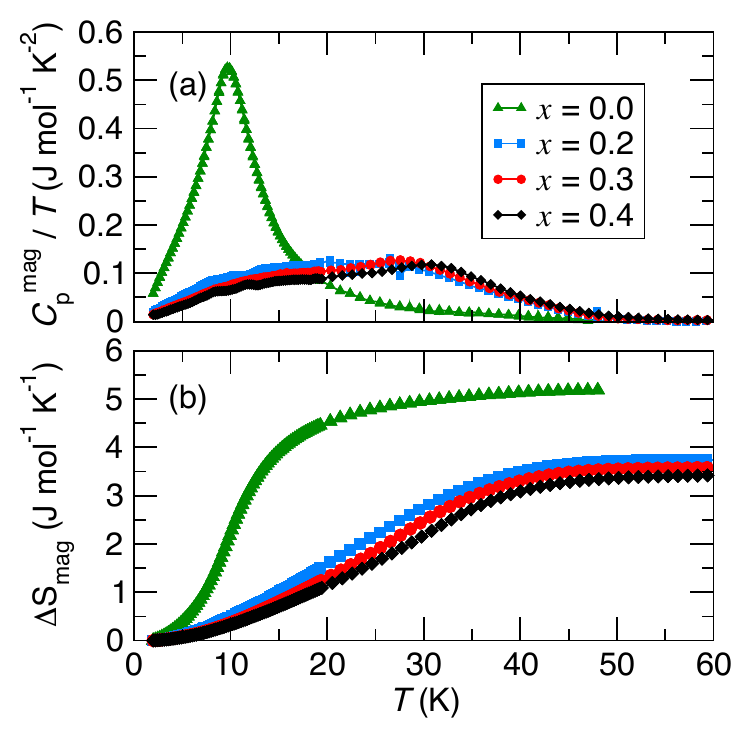}
\caption{\label{fig:hc} (Color online) (a) Temperature normalized magnetic heat capacity
[$(C_p~-~C_{\text{lattice}})/T$] as a function of temperature indicating a
broad hump centered around $T_c$ for all compounds.  (b) Entropy release
through the magnetic ordering transition illustrating the remainder of spin disorder 
down to 2 K and  large entropy losses above the magnetic transition temperature. }
\end{center}
\end{figure}

Further elucidation of the temperature dependent magnetism is provided with the 
magnetic specific heat for each compound, determined using non-magnetic analogs, 
 Mg(OH)$_2$ for $x=0$  and Zn(OH)$_{1.6}$(Cl)$_
{0.4}$(H$_2$O)$_{0.2}$ for $x>0$, to subtract contributions from the lattice.  To account for the different oscillator masses of the non-magnetic analogs, their temperature axes were scaled by a ratio of the calculated Debye temperatures ($\Theta_{\text{D}}$) as demonstrated in Ref.\,\onlinecite{Tari_2003}.  The temperature axis of the non-magnetic analog was divided by the ratio $\Theta_{\text{D}}^{non-mag}/\Theta_{\text{D}}^{magnetic}$.    The magnetic contribution to the specific heat capacity is then obtained from the algebraic subtraction of the host (non-magnetic) compound from the experimentally measured specific heat capacity, both illustrated in Fig.\,\ref{fig:hctotal}. 

The low temperature exponents ($T^\beta$, Table\,\ref{tab:summary}) resemble those of two-dimensional magnetic systems, indicating weak interlayer coupling.\cite{xymagnets}  There is a significant fraction of entropy lost above $T_c$, corollary to the observation of short-range spin interactions above the ordering temperature from the Curie-Weiss analysis (Fig.\,\ref{fig:normcw}).  The total entropy lost through the transition is only 25\% of the expected value for full ordering of a $S = 3/2$ system [$\Delta S=R\ln(2S+1)$] for $x>0$  (Fig.\,\ref{fig:hc}, Table\,\ref{tab:summary}).  Thus, spin disorder persists even at 2 K.  Using neutron diffraction, we identify any periodic magnetic order.

\subsection{Neutron Diffraction}

\begin{figure}
\includegraphics[width=3.25in]{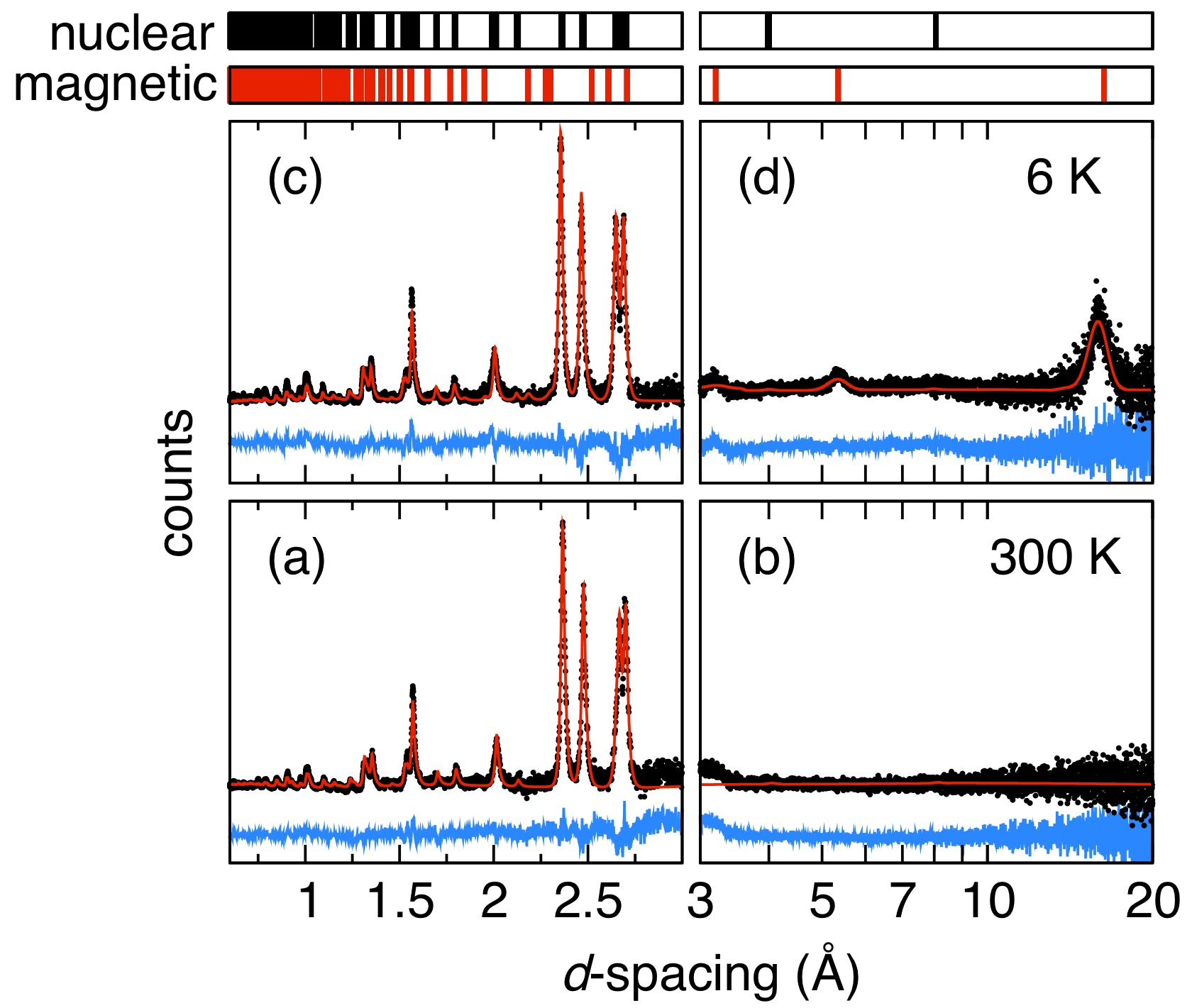}
\caption{\label{fig:neutron} Rietveld refinement of the nuclear and magnetic structures obtained from time-of-flight neutron powder diffraction from the HIPD instrument at (a, b) 300 K and (c, d) 6 K  for Co(OD)$_{1.6}$(Cl)$_{0.4}$(D$_2$O)$_{0.3}$ with (a, c) $90^\circ$ and (b, d) $14^\circ$ detector banks, showing the absence of magnetic reflections at large $d$-spacings and a significant contribution of diffuse nuclear scattering at $\sim$3 \AA.  Expected Bragg planes of the nuclear (black hashes) and magnetic [red hashes, $\mathbf{k}=(0,~0,~1.5)$] phases are shown above, with the Rietveld refined profiles (red lines) above the difference curves (blue lines).}
\end{figure}

Magnetic Bragg scattering from time-of-flight neutron powder diffraction of 
Co(OD)$_{1.6}$(Cl)$_{0.4}$(D$_2$O)$_{0.3}$ ($x=0.4$; Fig.\,\ref{fig:neutron}) indicates a N\'eel antiferromagnetic structure.  The average magnetic structure, illustrated in Fig.\,\ref{fig:structuremag}(a), describes a spin arrangement with in-plane ferromagnetically coupled spins on octahedral cobalt sites that are antiferromagnetically aligned to adjoining tetrahedral sites.  Each layer is antiferromagnetically coupled, yielding a fully compensated magnetic structure, inconsistent with the observed behavior.  

To arrive at this result, the unit-cell from Ref.\,\onlinecite{Neilson_ChemEurJ_2010} was used to initialize the refinement of the nuclear structure at 300\,K using \textsc{Fullprof}.\cite{fullprof}  The refinement was simultaneously carried out across multiple time-of-flight detector banks (14$^\circ$, 40$^\circ$, and 90$^\circ$) to achieve a broad range of momentum transfer and resolution, as shown in Figure\,\ref{fig:neutron}(a,b).  Deuterium positions were added approximately 1~\AA~from the $\mu$-(OD) positions and in the interlayer space neighboring the water oxygen position using an equivalent Wyckoff site with twice the occupancy.   Both deuterium positions were allowed to refine, along with all other atom positions.  The in-plane deuterium and oxygen positions ($x$,~$-x$,~0.5) are correlated and lead to divergent refinements when unconstrained; therefore, their positions were fixed.  The exceptionally high atomic displacement parameters (ADP) clearly indicate static disorder of the interlayer species and the deuterium bound to $\mu$-(OD) bridging units.  The structural parameters of the nuclear structure and refinement statistics are summarized in Table\,\ref{tbl:structureparameters}.  Significant diffuse scattering centered around 3 \AA ~($d$-spacing) is observed at both 300 K and 6 K and also indicates the presence of a large degree of static structural disorder, as previously studied.\cite{Neilson_ChemEurJ_2010}

The nuclear structure deduced from Rietveld refinement at 300 K was used to initialize the refinement at 6 K.  Upon cooling to 6 K, several broad Bragg reflections arise at $d=$ 3.20\,\AA, 5.35\,\AA, and 16.31\,\AA ~[Fig.\,\ref{fig:neutron}(c,d)], corresponding to a magnetic propagation vector of $\mathbf{k}=(0,~0,~1.5)$, commensurate with the nuclear unit cell [$R\bar{3}m$, $a = 3.136$, $c = 23.928$].  Examination of the profile backgrounds does not reveal any additional diffuse scattering at 6 K as evidence for short-range spin correlations.

\begin{table}
\begin{ruledtabular}
  \caption{Structure parameters of Co(OD)$_{1.6}$(Cl)$_{0.4}$(D$_2$O)$_{0.3}$ obtained from Rietveld refinement of data collected at 300 K from time-of-flight neutron powder diffraction, showing fractional coordinates ($x,y,z$), chemical site occupancies ($g$), and isotropic atomic displacement parameters ($B_{iso}$) for a rhombohedral unit cell in a hexagonal setting, $a = 3.14266(7)$, $c = 24.03738(2)$, $R\bar{3}m$ (166). $^a$ }
  \label{tbl:structureparameters}
    \begin{tabular}{c|c|c|c|c|c|c}
    Atom & Site & $x$ & $y$ & $z$ & $g$ & $B_{iso}$ \\
    \hline
    Co(1) & $3a$ & 0 & 0 &0 & 0.765 & 0.2(9) \\
   Co(2) & $6a$ & 0 & 0 & 0.069(5) & 0.235 & 0.2(9) \\
      Cl(1) & $6c$ & 0 & 0  & 0.172(2) & 0.235 & 5.2(9) \\
   O(1) & $6c$ & $1/3$ & $2/3$  & 0.045(1) & 1 & 2.5(7) \\
    O(2) & $18h$ & 0.07& 0.93 & 0.5 & 0.076(5) & 1.1(9) \\
      D(1) & $6c$ & $1/3$ & $2/3$ & 0.080(3)& 1 & 23(9) \\
   D(2) & $18h$ & $0.4$ & $0.6$ & 0.5& 0.15(1) & 44(9) \\
  \end{tabular}\\
   $^a$ Refinement statistics: 90$^\circ$ bank: $R_{p} = 4.1\%$, $R_{wp} = 5.6\%$, $\chi^2 = 4.2$; 40$^\circ$ bank: $R_{p} = 4.2\%$, $R_{wp} = 5.8\%$, $\chi^2 = 2.68$; 14$^\circ$ bank: $R_{p} = 3.4\%$, $R_{wp} = 4.8\%$, $\chi^2 = 9.1$.  
\end{ruledtabular}
\end{table}

Using \textsc{BasIreps},\cite{basireps} the irreducible representation (IR) of the Little Group, $G_{\mathbf{k}}$, was decomposed with two unique magnetic atoms (Co$^\text{oct}$ and Co$^\text{tet}$), as summarized in Table\,\ref{tbl:basisvector}.  The octahedrally coordinated site decomposes into two IR's ($\Gamma_1$ and $\Gamma_2$).  The tetrahedrally coordinated cobalt defines two unique atoms [Co$^\text{tet}$(1),  Co$^\text{tet}$(2)] in 4 IR's ($\Gamma_1$, $\Gamma_2$, $\Gamma_3$, and $\Gamma_4$).  We assume that both cobalt polyhedra belong to the same IR because of inter-site connectivity and strong exchange interactions.

Of the allowed IR's, only $\Gamma_2$ captures the intensity of the magnetic diffraction peaks.  In $\Gamma_2$, two basis vectors with real and imaginary components describe the moment contribution on each site.  However, powder and domain averaging precludes determination of a unique orientation of the basis vectors within the $ab$ plane of a hexagonal unit cell.   Therefore, only the Fourier coefficients ($c_\text{oct}$, $c_\text{tet}$) of the basis vector $\psi_1$ within $\Gamma_2$ for the Co$^{\text{oct}}$ and Co$^{\text{tet}}$ sites were allowed to refine.
While the cobalt positions are well-defined but partially occupied, refinement to the magnetic reflections reveals partial spin disorder.
The basis vector Fourier coefficients indicate reduced moments of both cobalt sites: 1.2 
$\mu_\text{B}$ for Co$^{\text{oct}}$, and 1.1 $\mu_\text{B}$ for Co$^{\text{tet}}$ ($R_{\text{mag}} = 14.8\%$, 14$^\circ$ detector bank) .  

In studies of dilute anisotropic antiferromagnets, magnetic Bragg reflections from quasi-elastic neutron scattering exhibit increased Lorenztian width from the nuclear reflections and reduced scattering intensity with applied fields.\cite{PhysRevB.28.1438, PhysRevLett.48.438, 
PhysRevB.28.2602}   In those random dilute systems, the formation of magnetic microdomains and the disruption of magnetic long-range order is driven by local random-field energy.\cite{PhysRevB.28.1438}  While a different experiment here, the magnetic reflections from neutron diffraction have significantly increased Lorenztian widths and low diffracted intensities, which also indicate disrupted long-range order of the magnetic structure.   Here, we hypothesize that uncompensated moments from the local clustering of polyhedra disrupts the long-range magnetic order into microdomains, and thus decreases the diffracted intensity and broadens the reflections.  

\begin{table}
\begin{ruledtabular}
\caption{\label{tbl:basisvector}
Basis vectors for the space group $R\bar{3}m$ in a hexagonal setting  ($a = 3.136$, $c = 23.928$) with the propagation vector ${\bf k}=( 0,~ 0,~ 1.5)$.  
The magnetic representation for  the Co$^{\text{oct}}$ site, $( 0,~ 0,~ 0)$ can be decomposed into the tabulated irreducible representations of $G_\mathbf{k}$.  The atoms of the nonprimitive basis for  Co$^{\text{tet}}$ are defined according to 1: $( 0,~ 0,~0.069)$, 2: $( 0,~ 0,-0.069)$.}
\begin{tabular}{ccc|cccccc}
  &   &  & \multicolumn{6}{c}{BV components} \\
   IR     &     BV    &       Atom      & $m_{\|a}$ & $m_{\|b}$ & $m_{\|c}$ & $im_{\|a}$ & $im_{\|b}$ & $im_{\|c}$ \\
\hline
$\Gamma_{1}$ & $\psi_{1}$ &  Co$^{\text{oct}}$  &      0 &      0 &     1  &0&0&0\\
$\Gamma_{2}$ & $\psi_{1}$ & Co$^{\text{oct}}$   &      1 &      0 &     0 & $-0.58$ & $-1.15$& 0  \\
                           & $\psi_{2}$ & Co$^{\text{oct}}$  &     0 &      1 &     0 & $-1.15$ &$-0.58$ & 0  \\
   \hline
$\Gamma_{1}$  & $\psi_{1}$ &      Co$^{\text{tet}}$(1) &      0 &      0 &      1  &0&0&0\\
             &              &                             Co$^{\text{tet}}$(2) &      0 &      0 &      1  &0&0&0\\
$\Gamma_{2}$  & $\psi_{1}$ &       Co$^{\text{tet}}$(1)&     1.5 &      0 &      0   &$-0.87$&$-1.73$&0\\
             &              &                              Co$^{\text{tet}}$(2) &    1.5 &      0 &      0   &$-0.87$&$-1.73$&0\\
             &                 $\psi_{2}$ &       Co$^{\text{tet}}$(1) &       0 &    1.5 &      0  &$-1.73$ & $-0.87 $& 0\\
             &              &                              Co$^{\text{tet}}$(2)&     0 &    1.5 &      0  &$-1.73$ & $-0.87$ & 0 \\            
$\Gamma_{3}$  & $\psi_{1}$ &     Co$^{\text{tet}}$(1) &      0 &      0 &      1   &0&0&0\\
             &              &                             Co$^{\text{tet}}$(2) &      0 &      0 &      $-1$  &0&0&0\\
$\Gamma_{4}$  & $\psi_{1}$ &      Co$^{\text{tet}}$(1) &      1.5 &      0 &      0  &$-0.87$ &$-1.73$&0\\
             &              &                             Co$^{\text{tet}}$(2) &     $-1.5$ &      0 &      0  &0.87&1.73&0 \\
             &                  $\psi_{2}$ &      Co$^{\text{tet}}$(1) & 0 &$ -1.5$ &      0  &1.73&0.87&0\\
             &              &                              Co$^{\text{tet}}$(2) &  0 & 1.5 &      0 &$-1.73$&$-0.87$&0\\
         
\end{tabular}
\end{ruledtabular}
\end{table}

\subsection{Magnetism from Local Structure}

\begin{figure}
\centering
\includegraphics[width =2.5in]{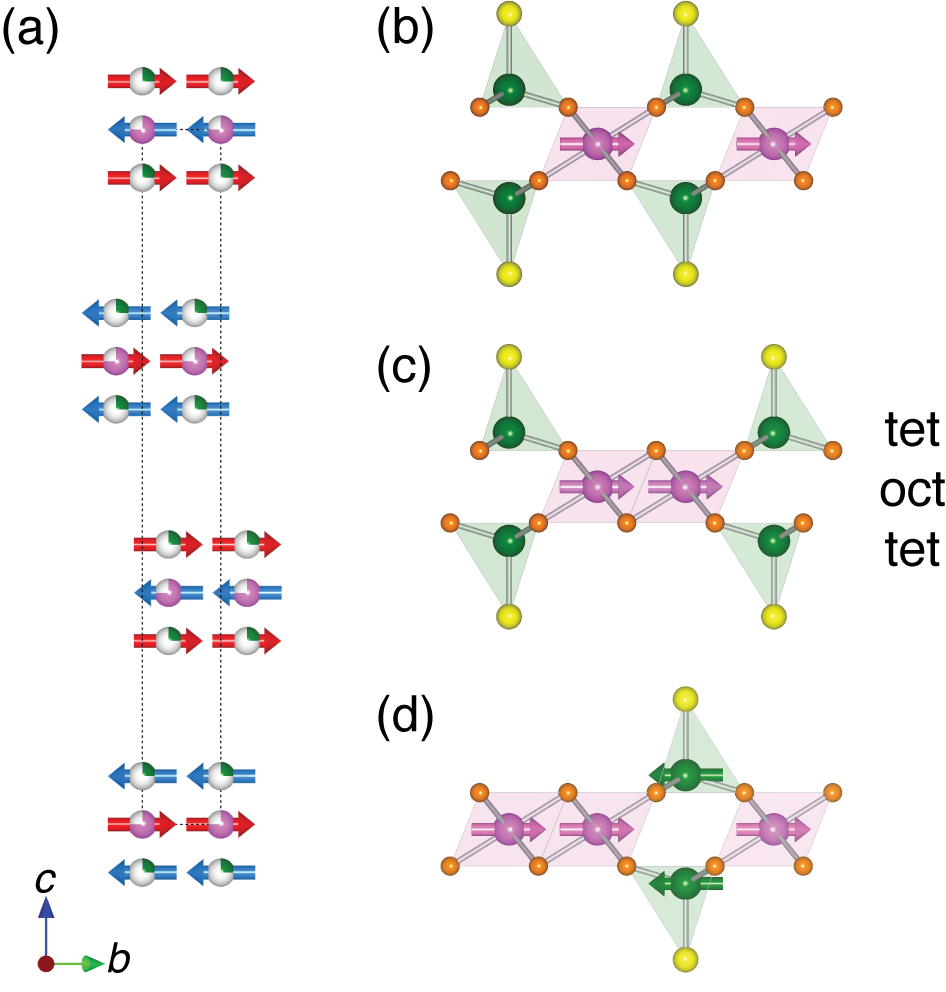}
\caption{\label{fig:structuremag} (Color online) (a) Symmetry allowed spin configuration of
Co(OD)$_{1.6}$(Cl)$_{0.4}$(D$_2$O)$_{0.3}$ at 6\,K (dashed 
line indicates the nuclear unit-cell).  (b-d) Locally ordered motifs (b) [Co$^{\text{oct}}_3$Co$^{\text{tet}}_2$], (c) [Co$^{\text{oct}}_8$Co$^{\text{tet}}_2$], and (d) [Co$^{\text{oct}}_3$Co$^{\text{tet}}_2$] illustrate possible microscopic spin configurations compatible with the average magnetic structure and bulk magnetization.  Pink spheres denote Co$^{\text{oct}}$, green Co$^{\text{tet}}$; oxygen is orange, chlorine yellow, and deuterium is omitted for clarity.   }
\end{figure}

To reconcile ambiguities engendered from fractional occupancy in the unit-cell, we refer to our previous local structure study of the pair distribution functions of layered cobalt hydroxide chlorides,\cite{Neilson_ChemEurJ_2010} depicted in Fig.\,\ref{fig:structure}(b-d).  We represent the structure from a distribution of [Co$^{\text{oct}}_{n}$Co$^{\text{tet}}_2$] motifs with a set number of octahedra between tetrahedra.  These randomly distributed motifs serve as the microdomains for describing the observed magnetic behavior, redrawn in Fig.\,\ref{fig:structuremag}.  

Microscopically, [Co$^{\text{oct}}_{15}$Co$^{\text{tet}}_2$] motifs [Fig.\,\ref{fig:structuremag}(d)]  appear as N\'eel domains, with ferromagnetically coupled octahedral sites, antiparallel to tetrahedral sites, to yield a net local moment.  In the [Co$^{\text{oct}}_3$Co$^{\text{tet}}_2$] and [Co$^{\text{oct}}_8$Co$^{\text{tet}}_2$] domains, illustrated in Fig.\,\ref{fig:structuremag}(b,c), the edge-sharing octahedral sites are ferromagnetically coupled within the layer.  However,  the tetrahedral sites likely remain disordered, as explained below.    
 
 The presence of defect spins, as from tetrahedrally coordinated cobalt sites, often leads to spin disorder.  For example, defect spins placed adjacent to well-ordered ferromagnetic layers, as seen in $(1-x)$FeTiO$_3$-$x$Fe$_2$O$_3$ solid solutions,\cite{Ishikawa_JPSJ_1962, Ishikawa_JPSJ_1985, Arai_JPSJ_1985} results in cluster spin glass behavior.   Here, the strong magnetocrystalline anisotropy of octahedral Co$^{2+}$ prevents the canting of spins from the crystal field axes, giving them an Ising character.\cite{PTP.17.197}  In the case of high defect concentrations, such as in [Co$^{\text{oct}}_3$Co$^{\text{tet}}_2$] and [Co$^{\text{oct}}_8$Co$^{\text{tet}}_2$] domains, competition between the dominant ferromagnetically coupled layers and the magnetocrystalline anisotropy prevents the formation of a well-ordered ground state.  This is supported by the observed trend in decreasing magnetic entropy release with increasing Co$^{\text{tet}}$ sites (Table\,\ref{tab:summary}).   

While Goodenough-Kanamori-Anderson superexchange considerations suggest a preference for antiferromagnetic coupling between distinct polyhedra, this interaction is weak, as indicated from the sign change and decreasing $|\Theta_\text{CW}|/T_c$ with increasing $x$.   On average, a small fraction of tetrahedral spins must align antiparallel to the octahedra, as determined from magnetic neutron diffraction.  Disordered tetrahedral spins disrupt the long-range order of interlamellar dipolar coupling, permitting a low field spin-flop transition.  This is observed in the initial magnetization by a change in curvature in $M(H)$ at $\sim$0.4 T (Fig.\,\ref{fig:hys}).  As the field is reversed, neighboring N\'eel ordered Co$^{\text{tet}}$ spins from distinct layers couple \emph{via} dipolar interactions along the interlayer space, preventing immediate reversal of the field induced transition.  This gives rise to a remanent magnetization (Fig.\,\ref{fig:hys}) and a large frequency dependent amplitude and phase in $\chi(\omega,T)$ (Supplemental Information).  While the disparity between the observed magnetization and the fully compensated average N\'eel antiferromagnetic structure can be reconciled with a spin flop transition, this cannot account for the nearly composition independent constant  saturation magnetization, $T_c$, and entropy release.  Therefore, a local analysis of the structure must be applied to reconcile these differences.

\begin{table}
\caption{\label{tab:local}Summary of local atomic phase fractions from Ref. \onlinecite
{Neilson_ChemEurJ_2010}.}  
\begin{ruledtabular}
\begin{tabular}{cccc}
 &[Co$^{\text{oct}}_3$Co$^{\text{tet}}_2$]&[Co$^{\text{oct}}_8$Co$^{\text{tet}}_2$]&[Co$^{\text{oct}}_{15}$Co$^{\text{tet}}_2$]\\
  $x$&$A$&$B$&$C$\\
\hline
0.2 & 22\%&47\%&31\% \\
0.3 &29\%&51\%& 21\% \\
0.4  &36\%&49\%&15\%\\
\end{tabular}
\end{ruledtabular}
\label{tbl:local}
\end{table}%

From the polyhedral composition provided by the real-space analysis,\cite
{Neilson_ChemEurJ_2010} we calculate the individual moment contributions from 
each local motif with the assumption that all Co$^{\text{oct}}$ are parallel, and dilute Co$^{\text{tet}}$ spins (as in [Co$^{\text{oct}}_{15}$Co$^{\text{tet}}_2$]) are antiparallel to Co$^{\text{oct}}$ in a N\'eel configuration  [Fig.\,\ref{fig:structuremag}(b-d)].  
Letting $A$, $B$, and $C$ represent the atomic phase fractions (Table\,\ref{tab:local})  of the local motifs provided from Ref.~\onlinecite
{Neilson_ChemEurJ_2010} for [Co$^{\text{oct}}_3$Co$^{\text{tet}}_2$], [Co$^{\text{oct}}_8$Co$^{\text{tet}}_2$], [Co$^{\text{oct}}_{15}$Co$^{\text{tet}}_2$], respectively, the summed moment from each 
microdomain is calculated as:
\begin{equation}
M_{sat}= \frac{3}{2} \left( \frac{3A + 8B + (15-2)C}{5A + 10B + 17C}\right) \label
{eqn:count}
\end{equation}
Using Eqn.\,\ref{eqn:count}, the locally derived saturation moments for $x$~= 0.2, 0.3, 
and 0.4 are tabulated in Table \ref{tab:summary} (Local $M_{sat}$).  
We compute a negligible dependence of $M_{sat}$ on composition with values comparable 
to those observed.  
The agreement of the observed properties with the assembly of local contributions 
reconciles the inaccuracies resulting from the disordered average structure and 
microdomain-like magnetic diffraction.  

\section{Conclusion}

While the average symmetry-allowed magnetic structure of the layered cobalt hydroxides appears as a fully 
compensated N\'eel-type antiferromagnet, the magnetic behavior is better explained by 
local and randomly distributed intralayer N\'eel ferrimagnetic regions that can be aligned with 
an applied field, interspersed with disordered defect spins that interrupt compensation 
of the ferrimagnetic regions from layer to layer.  
We present an extreme case in which magnetic site occupancy exhibits no long-range 
order, yet a long-range ordered magnetic structure is observed, on average.  
Analyses using probes of average structural and magnetic structures are typically interpreted under the assumption that a single unit-cell suffices to describe systems in their entirety.  This assumption breaks down when there is disorder.  
We illustrate here that the local atomic structure offers an alternative structural 
mechanism for introducing random fields in a magnetic lattice to yield microdomain 
behavior.

\section{Acknowledgements}

JRN gratefully thanks the NSF for support through the Graduate Research Fellowship.  The 
authors thank A. Llobet, E.E.\,Rodriguez and T.M.\,McQueen for helpful discussions.  DPS acknowledges support from the UCSB-LANL Institute for Multiscale Materials Studies.  JAK thanks the ConvEne-IGERT Program (NSF-DGE 0801627) for an Associateship. This research was supported by a grant from the DOE, Office of Basic Energy Science to DEM (DEFG03-02ER46006) and to JAK (DE-FG02-10ER16081); and by use of the time-of-flight beam line HIPD at DOE's Lujan Center at  Los Alamos National Laboratory (operated by Los Alamos National Security, LLC under DOE under Contract No. DE-AC52-06NA25396), and UCSB's Materials Research Laboratory (MRL) Central Facilities, supported by the MRSEC Program (NSF, Award No. DMR05-20415), to which BCM also acknowledges financial support; the MRL is a member of the NSF-funded Materials Research Facilities Network (www.mrfn.org).


\begin{thebibliography}{23}
\expandafter\ifx\csname natexlab\endcsname\relax\def\natexlab#1{#1}\fi
\expandafter\ifx\csname bibnamefont\endcsname\relax
  \def\bibnamefont#1{#1}\fi
\expandafter\ifx\csname bibfnamefont\endcsname\relax
  \def\bibfnamefont#1{#1}\fi
\expandafter\ifx\csname citenamefont\endcsname\relax
  \def\citenamefont#1{#1}\fi
\expandafter\ifx\csname url\endcsname\relax
  \def\url#1{\texttt{#1}}\fi
\expandafter\ifx\csname urlprefix\endcsname\relax\def\urlprefix{URL }\fi
\providecommand{\bibinfo}[2]{#2}
\providecommand{\eprint}[2][]{\url{#2}}

\bibitem[{\citenamefont{Ramirez}(1994)}]{Ramirez_GeoFrust}
\bibinfo{author}{\bibfnamefont{A.~P.} \bibnamefont{Ramirez}},
  \bibinfo{journal}{Annu. Rev. Mater. Sci.} \textbf{\bibinfo{volume}{24}},
  \bibinfo{pages}{453} (\bibinfo{year}{1994}).

\bibitem[{\citenamefont{Fishman and Aharony}(1979)}]{Fishman_1979}
\bibinfo{author}{\bibfnamefont{S.}~\bibnamefont{Fishman}} \bibnamefont{and}
  \bibinfo{author}{\bibfnamefont{A.}~\bibnamefont{Aharony}},
  \bibinfo{journal}{J.Phys. C} \textbf{\bibinfo{volume}{12}},
  \bibinfo{pages}{L729} (\bibinfo{year}{1979}).

\bibitem[{\citenamefont{Yoshizawa et~al.}(1982)\citenamefont{Yoshizawa, Cowley,
  Shirane, Birgeneau, Guggenheim, and Ikeda}}]{PhysRevLett.48.438}
\bibinfo{author}{\bibfnamefont{H.}~\bibnamefont{Yoshizawa}},
  \bibinfo{author}{\bibfnamefont{R.~A.} \bibnamefont{Cowley}},
  \bibinfo{author}{\bibfnamefont{G.}~\bibnamefont{Shirane}},
  \bibinfo{author}{\bibfnamefont{R.~J.} \bibnamefont{Birgeneau}},
  \bibinfo{author}{\bibfnamefont{H.~J.} \bibnamefont{Guggenheim}},
  \bibnamefont{and} \bibinfo{author}{\bibfnamefont{H.}~\bibnamefont{Ikeda}},
  \bibinfo{journal}{Phys. Rev. Lett.} \textbf{\bibinfo{volume}{48}},
  \bibinfo{pages}{438} (\bibinfo{year}{1982}).

\bibitem[{\citenamefont{Hagen et~al.}(1983)\citenamefont{Hagen, Cowley, Satija,
  Yoshizawa, Shirane, Birgeneau, and Guggenheim}}]{PhysRevB.28.2602}
\bibinfo{author}{\bibfnamefont{M.}~\bibnamefont{Hagen}},
  \bibinfo{author}{\bibfnamefont{R.~A.} \bibnamefont{Cowley}},
  \bibinfo{author}{\bibfnamefont{S.~K.} \bibnamefont{Satija}},
  \bibinfo{author}{\bibfnamefont{H.}~\bibnamefont{Yoshizawa}},
  \bibinfo{author}{\bibfnamefont{G.}~\bibnamefont{Shirane}},
  \bibinfo{author}{\bibfnamefont{R.~J.} \bibnamefont{Birgeneau}},
  \bibnamefont{and} \bibinfo{author}{\bibfnamefont{H.~J.}
  \bibnamefont{Guggenheim}}, \bibinfo{journal}{Phys. Rev. B}
  \textbf{\bibinfo{volume}{28}}, \bibinfo{pages}{2602} (\bibinfo{year}{1983}).

\bibitem[{\citenamefont{Neilson et~al.}(2010)\citenamefont{Neilson, Kurzman,
  Seshadri, and Morse}}]{Neilson_ChemEurJ_2010}
\bibinfo{author}{\bibfnamefont{J.~R.} \bibnamefont{Neilson}},
  \bibinfo{author}{\bibfnamefont{J.~A.} \bibnamefont{Kurzman}},
  \bibinfo{author}{\bibfnamefont{R.}~\bibnamefont{Seshadri}}, \bibnamefont{and}
  \bibinfo{author}{\bibfnamefont{D.~E.} \bibnamefont{Morse}},
  \bibinfo{journal}{Chem. Eur. J.} \textbf{\bibinfo{volume}{16}},
  \bibinfo{pages}{9998} (\bibinfo{year}{2010}).

\bibitem[{\citenamefont{Takada et~al.}(1966)\citenamefont{Takada, Bando,
  Kiyama, and Miyamoto}}]{Takada_1966}
\bibinfo{author}{\bibfnamefont{T.}~\bibnamefont{Takada}},
  \bibinfo{author}{\bibfnamefont{Y.}~\bibnamefont{Bando}},
  \bibinfo{author}{\bibfnamefont{M.}~\bibnamefont{Kiyama}}, \bibnamefont{and}
  \bibinfo{author}{\bibfnamefont{H.}~\bibnamefont{Miyamoto}},
  \bibinfo{journal}{J. Phys. Soc. Japan} \textbf{\bibinfo{volume}{21}},
  \bibinfo{pages}{2726} (\bibinfo{year}{1966}).

\bibitem[{\citenamefont{Liu et~al.}(2008)\citenamefont{Liu, Liu, Hu, Guo, Lv,
  Cui, Zhao, and Zhang}}]{Liu_APL_2008}
\bibinfo{author}{\bibfnamefont{X.~H.} \bibnamefont{Liu}},
  \bibinfo{author}{\bibfnamefont{W.}~\bibnamefont{Liu}},
  \bibinfo{author}{\bibfnamefont{W.~J.} \bibnamefont{Hu}},
  \bibinfo{author}{\bibfnamefont{S.}~\bibnamefont{Guo}},
  \bibinfo{author}{\bibfnamefont{X.~K.} \bibnamefont{Lv}},
  \bibinfo{author}{\bibfnamefont{W.~B.} \bibnamefont{Cui}},
  \bibinfo{author}{\bibfnamefont{X.~G.} \bibnamefont{Zhao}}, \bibnamefont{and}
  \bibinfo{author}{\bibfnamefont{Z.~D.} \bibnamefont{Zhang}},
  \bibinfo{journal}{Appl. Phys. Lett.} \textbf{\bibinfo{volume}{93}},
  \bibinfo{pages}{202502} (\bibinfo{year}{2008}).

\bibitem[{\citenamefont{Oswald and Asper}(1977)}]{Oswald_Asper_1977}
\bibinfo{author}{\bibfnamefont{H.~R.} \bibnamefont{Oswald}} \bibnamefont{and}
  \bibinfo{author}{\bibfnamefont{R.}~\bibnamefont{Asper}},
  \emph{\bibinfo{title}{Preparation and Crystal Growth of Materials with
  Layered Structures}} (\bibinfo{publisher}{D. Riedel Publishing},
  \bibinfo{year}{1977}), chap. \bibinfo{chapter}{Bivalent Metal Hydroxides}.

\bibitem[{\citenamefont{Neilson et~al.}(2009)\citenamefont{Neilson, Schwenzer,
  Seshadri, and Morse}}]{Neilson_IC_2009}
\bibinfo{author}{\bibfnamefont{J.~R.} \bibnamefont{Neilson}},
  \bibinfo{author}{\bibfnamefont{B.}~\bibnamefont{Schwenzer}},
  \bibinfo{author}{\bibfnamefont{R.}~\bibnamefont{Seshadri}}, \bibnamefont{and}
  \bibinfo{author}{\bibfnamefont{D.~E.} \bibnamefont{Morse}},
  \bibinfo{journal}{Inorg. Chem.} \textbf{\bibinfo{volume}{48}},
  \bibinfo{pages}{11017} (\bibinfo{year}{2009}).

\bibitem[{\citenamefont{Peterson et~al.}(2000)\citenamefont{Peterson, Gutmann,
  Proffen, and Billinge}}]{PDFgetN}
\bibinfo{author}{\bibfnamefont{P.~F.} \bibnamefont{Peterson}},
  \bibinfo{author}{\bibfnamefont{M.}~\bibnamefont{Gutmann}},
  \bibinfo{author}{\bibfnamefont{T.}~\bibnamefont{Proffen}}, \bibnamefont{and}
  \bibinfo{author}{\bibfnamefont{S.~J.~L.} \bibnamefont{Billinge}},
  \bibinfo{journal}{J. Appl. Crystallogr.} \textbf{\bibinfo{volume}{33}},
  \bibinfo{pages}{1192} (\bibinfo{year}{2000}).

\bibitem[{\citenamefont{Melot et~al.}(2009)\citenamefont{Melot, Drewes,
  Seshadri, Stoudenmire, and Ramirez}}]{Melot_2009_JPCM}
\bibinfo{author}{\bibfnamefont{B.~C.} \bibnamefont{Melot}},
  \bibinfo{author}{\bibfnamefont{J.~E.} \bibnamefont{Drewes}},
  \bibinfo{author}{\bibfnamefont{R.}~\bibnamefont{Seshadri}},
  \bibinfo{author}{\bibfnamefont{E.~M.} \bibnamefont{Stoudenmire}},
  \bibnamefont{and} \bibinfo{author}{\bibfnamefont{A.~P.}
  \bibnamefont{Ramirez}}, \bibinfo{journal}{J. Phys.: Condens. Matter}
  \textbf{\bibinfo{volume}{21}}, \bibinfo{pages}{216007}
  (\bibinfo{year}{2009}).

\bibitem[{\citenamefont{Rabu et~al.}(1993)\citenamefont{Rabu, Angelov, Legoll,
  Belaiche, and Drillon}}]{Rabu_IC_1993}
\bibinfo{author}{\bibfnamefont{P.}~\bibnamefont{Rabu}},
  \bibinfo{author}{\bibfnamefont{S.}~\bibnamefont{Angelov}},
  \bibinfo{author}{\bibfnamefont{P.}~\bibnamefont{Legoll}},
  \bibinfo{author}{\bibfnamefont{M.}~\bibnamefont{Belaiche}}, \bibnamefont{and}
  \bibinfo{author}{\bibfnamefont{M.}~\bibnamefont{Drillon}},
  \bibinfo{journal}{Inorg. Chem.} \textbf{\bibinfo{volume}{32}},
  \bibinfo{pages}{2463} (\bibinfo{year}{1993}).

\bibitem[{\citenamefont{Kurmoo}(1990)}]{Kurmoo_1999}
\bibinfo{author}{\bibfnamefont{M.}~\bibnamefont{Kurmoo}},
  \bibinfo{journal}{Phil. Trans. R. Soc. Lond. A}
  \textbf{\bibinfo{volume}{357}}, \bibinfo{pages}{3041} (\bibinfo{year}{1990}).

\bibitem[{\citenamefont{Kurmoo}(1999)}]{Kurmoo:1999uq}
\bibinfo{author}{\bibfnamefont{M.}~\bibnamefont{Kurmoo}},
  \bibinfo{journal}{Chem. Mater.} \textbf{\bibinfo{volume}{11}},
  \bibinfo{pages}{3370} (\bibinfo{year}{1999}).

\bibitem[{\citenamefont{Tari}(2003)}]{Tari_2003}
\bibinfo{author}{\bibfnamefont{A.}~\bibnamefont{Tari}},
  \emph{\bibinfo{title}{The Specific Heat of Matter at Low Temperatures}}
  (\bibinfo{publisher}{Imperial College Press}, \bibinfo{year}{2003}).

\bibitem[{\citenamefont{Regnault and Rossat-Mignod}(1990)}]{xymagnets}
\bibinfo{author}{\bibfnamefont{L.~P.} \bibnamefont{Regnault}} \bibnamefont{and}
  \bibinfo{author}{\bibfnamefont{J.}~\bibnamefont{Rossat-Mignod}}, in
  \emph{\bibinfo{booktitle}{Magnetic Propertyies of Layered Transition Metal
  Compounds}}, edited by \bibinfo{editor}{\bibfnamefont{L.~J.}
  \bibnamefont{de~Jongh}} (\bibinfo{publisher}{Kluwer Academic Publishers},
  \bibinfo{year}{1990}).

\bibitem[{\citenamefont{Rodriguez-Carvajal}(1993)}]{fullprof}
\bibinfo{author}{\bibfnamefont{J.}~\bibnamefont{Rodriguez-Carvajal}},
  \bibinfo{journal}{Physica B} \textbf{\bibinfo{volume}{192}},
  \bibinfo{pages}{55} (\bibinfo{year}{1993}).

\bibitem[{\citenamefont{Rodriguez-Carvajal}(2001)}]{basireps}
\bibinfo{author}{\bibfnamefont{J.}~\bibnamefont{Rodriguez-Carvajal}},
  \bibinfo{journal}{Mater. Sci. Forum} \textbf{\bibinfo{volume}{378-381}},
  \bibinfo{pages}{268} (\bibinfo{year}{2001}).

\bibitem[{\citenamefont{Birgeneau et~al.}(1983)\citenamefont{Birgeneau,
  Yoshizawa, Cowley, Shirane, and Ikeda}}]{PhysRevB.28.1438}
\bibinfo{author}{\bibfnamefont{R.~J.} \bibnamefont{Birgeneau}},
  \bibinfo{author}{\bibfnamefont{H.}~\bibnamefont{Yoshizawa}},
  \bibinfo{author}{\bibfnamefont{R.~A.} \bibnamefont{Cowley}},
  \bibinfo{author}{\bibfnamefont{G.}~\bibnamefont{Shirane}}, \bibnamefont{and}
  \bibinfo{author}{\bibfnamefont{H.}~\bibnamefont{Ikeda}},
  \bibinfo{journal}{Phys. Rev. B} \textbf{\bibinfo{volume}{28}},
  \bibinfo{pages}{1438} (\bibinfo{year}{1983}).

\bibitem[{\citenamefont{Ishikawa}(1962)}]{Ishikawa_JPSJ_1962}
\bibinfo{author}{\bibfnamefont{Y.}~\bibnamefont{Ishikawa}},
  \bibinfo{journal}{J. Phys. Soc. Jpn.} \textbf{\bibinfo{volume}{17}},
  \bibinfo{pages}{1835} (\bibinfo{year}{1962}).

\bibitem[{\citenamefont{Ishikawa et~al.}(1985)\citenamefont{Ishikawa, Saito,
  Arai, Watanabe, and Takei}}]{Ishikawa_JPSJ_1985}
\bibinfo{author}{\bibfnamefont{Y.}~\bibnamefont{Ishikawa}},
  \bibinfo{author}{\bibfnamefont{N.}~\bibnamefont{Saito}},
  \bibinfo{author}{\bibfnamefont{M.}~\bibnamefont{Arai}},
  \bibinfo{author}{\bibfnamefont{Y.}~\bibnamefont{Watanabe}}, \bibnamefont{and}
  \bibinfo{author}{\bibfnamefont{H.}~\bibnamefont{Takei}}, \bibinfo{journal}{J.
  Phys. Soc. Jpn.} \textbf{\bibinfo{volume}{54}}, \bibinfo{pages}{312}
  (\bibinfo{year}{1985}).

\bibitem[{\citenamefont{Arai et~al.}(1985)\citenamefont{Arai, Ishikawa, Saito,
  and Takei}}]{Arai_JPSJ_1985}
\bibinfo{author}{\bibfnamefont{M.}~\bibnamefont{Arai}},
  \bibinfo{author}{\bibfnamefont{Y.}~\bibnamefont{Ishikawa}},
  \bibinfo{author}{\bibfnamefont{N.}~\bibnamefont{Saito}}, \bibnamefont{and}
  \bibinfo{author}{\bibfnamefont{H.}~\bibnamefont{Takei}}, \bibinfo{journal}{J.
  Phys. Soc. Jpn.} \textbf{\bibinfo{volume}{54}}, \bibinfo{pages}{781}
  (\bibinfo{year}{1985}).

\bibitem[{\citenamefont{Kanamori}(1957)}]{PTP.17.197}
\bibinfo{author}{\bibfnamefont{J.}~\bibnamefont{Kanamori}},
  \bibinfo{journal}{Prog. Theor. Phys.} \textbf{\bibinfo{volume}{17}},
  \bibinfo{pages}{197} (\bibinfo{year}{1957}).

\end{thebibliography}
\end{document}